\documentstyle[aps,prb,preprint,12pt]{revtex}

\begin{document}

\author{ Stability of correlated electronic systems under the 
influence of the electron-phonon interaction}

\author{
A.Greco and A. Dobry}

\address{
Facultad de Ciencias Exactas, Ingenier\'{\i}a y Agrimensura (UNR)\\
 and Instituto de F\'{\i}sica Rosario.\\
 Av. Pellegrini 250, 2000 Rosario, Argentina.} 
\maketitle
\vskip 0.5cm
\begin{abstract}
We have used an exact diagonalization technique to study
the stability of the $t-J$-Holstein and
Hubbard-Holstein models under the influence of the electron-phonon
interaction. Previous results have been obtained using
frozen-phonon technique or introducing only a few dynamical phonon
modes due to the large Hilbert space.
To check these results we have done exact diagonalization
in a small cluster (four sites) including all the phonon
modes allowed by symmetry. We compare our results with those obtained
by using the adiabatic approximation.
\end{abstract}

\newpage

\section{Introduction}

Strong correlation and electron-phonon interaction are two important
features of High Tc (HTC) Superconductors. This situation motivates
the study of electron-phonon (e-ph) interaction in strongly correlated
electronic systems. One important question is how stable the
system is in the presence of the e-ph interaction. If the system has a
robust stability, it is possible to use BCS-like theories, but if the
system has a tendency to instability, polaronic and bipolaronic \cite
{alexandrov} theories play an important role.

Many analitical \cite{kabanov} and numerical
\cite{alemanes,nosotros1,Zhong,Scalapino} 
works have been done to answer this question. Most
of the numerical works use the frozen-phonon (adiabatic)
approximation \cite{alemanes,nosotros1,Zhong} and neglect
dynamical effects.

In a previous paper we studied the role of dynamical phonons in
the $t-J$ model with clusters of 10 and 16 sites including the
phonon Hilbert space \cite{nosotros2}.
We found an instability to a charge density wave (CDW)
for moderate e-ph interactions.
However, due to the large size of the Hilbert space, only a few
phonon modes were included in the calculation. In this case
the sensibility of our results with respect to the number of phonon modes
is a matter of debate \cite{Poilblanc}.

In order to check the validity
of the truncation of the Hilbert space in the number of
phonon modes, we consider,
as in other works \cite{kab}, a small cluster (four sites)
including all the phonon modes allowed by symmetry.
In section II we present models, approximations and results;
while in section III we discuss the main conclusions of this study.

\section{Models and results}

The Hamiltonian of the $t-J$-Holstein ($t-J$-H) model is:

\begin{eqnarray}
H=-t\sum_{\scriptstyle {i,j,\sigma}} (\tilde c_{i,\sigma}^{\dagger}
\tilde
c_{j,\sigma}
+H.c.)+J \sum_{\scriptstyle{i,j}}(S_i S_j - n_i n_j/4) + \nonumber \\
\sum_{\scriptstyle{i}}(\frac{P_i^2}{2M}+K/2 u_i^2-\alpha u_i n_i)
\end{eqnarray}

\noindent where the first two terms correspond to the usual $t-J$ model, $t$
is the hopping term and $J$ is the antiferromagnetic exchange interaction. $%
\tilde c$ and $\tilde c^{\dagger }$ are the creation and destruction
operators of a hole in a simple occupied Hilbert space, respectively.
The following two term are the kinetic and elastic energy of ions
with mass $M$, and elastic force constant $K$. Finally, the last term is
the Holstein-type interaction between electrons and ions with strain
$\alpha $. In this interaction the movement of the internal molecular
degrees of freedom ($u$) affect the electron density at a given site.
This kind of interaction is important in HTC superconductors \cite{zeyher}.
For the case of cuprates, this internal degree of freedom simulates the
breathing mode, where the oxygens of the unit cell move toward
the Cu ion \cite{Picho}.
Hereafter, we are going to study the dynamics of one hole
in a 4-sites cluster.
\vskip 0.3cm
\noindent{\bf a)Adiabatic phonon calculation }
\vskip 0.3cm
\noindent In this approximation the kinetic energy of the ions
is neglegible with respect to the other terms in the Hamiltonian. This
approximation is exact when the mass of the ions are infinite.

We have a quantum problem with a classical variable $u$. We minimize,
iterativelly, the adiabatic potential $E_e[\{u\}]$ starting from an initial
displacement configuration $\{u\}$. In each step of the iteration,
we evaluate the ground state of the electronic system and the site
electron density $<n_i>$.
Using $<n_i>$ and the relation $u_i^{new}=-\frac \alpha K<n_i>$,
a new displacement set is obtained;
if it falls within a predetermined threshold from the
initial, we stop the calculation, otherwise we iterate the procedure.

We define the dimensionless electron-phonon constant $%
\lambda ={\alpha }^2/Kt$. With this definition and rescalling $\alpha
u_i\rightarrow u_i$ all the results can be writen in terms of $\lambda $ for
a given set of electronic parameters $t$ and $J$. We choose $t$ as unit of
energy.

We show in Fig.1 the equilibrium displacement of
a given site as a function of $\lambda$, for different values of $J$.
In order to describe our results we
choose the most representative site displacement.
We can distinguish two different regimes: $0.0<J<J_N$ and $%
J>J_N$, where $J_N=0.28$ for our cluster \cite{elbio}.

For $J$ smaller than $J_N$ the electronic model has a ferromagnetic ground
state (Nagaoka regime). The hole moves as a free spinless fermion. Therefore,
like in the Holstein model \cite{kabanov}, we expect a transition towards a
localized state at a critical value of $\lambda $( $\lambda_c$).
  It is possible to see in Fig.1
that the transition occurs for $\lambda \sim 2.2$ and this value is
practically independent of $J$. For $\lambda <2.2$ the system presents a
homogeneous phase, while for $\lambda >2.2$ an abrupt transition towards a
localized state occurs. 
In this state the hole is principally located in a given 
site and the equilibrium 
displacement associated with this site is larger than the other ones.
In the figure we show the biggest displacement $u_i$.
There is also an increase of the hole occupation in this site, which
shows the localization of the hole.
This phase was previously interpreted  as a ''polaronic''
state of the system\cite{kabanov,alemanes}.
 We call this state ''polaronic'' even, at this
stage of the calculation, the adiabatic condition is not broken down and a
true polaron formation implies a mixing between the electronic and phononic
degrees of freedom.

For $J>J_N$ the transition towards a localized phase occurs at smaller $%
\lambda $, due to the preexisting self-localization of the hole in
the antiferromagnetic background and the resulting band narrowing \cite
{alemanes,Zhong} . We can also see that the transition is smoother, and
before going to a ''polaronic'' state the system undergoes a breathing
type of deformation. This kind of phase is due to the nontrivial 
momentum value of the ground state of the system for $\lambda =0.0$ \cite
{elbio}.
\vskip 0.3cm
\noindent{\bf b)Full dynamics phonon calculation}
\vskip 0.3cm
\noindent In the momentum space for the phononic degrees of freedom,
the Hamiltonian (1) is:

\begin{eqnarray}
H=-t\sum_{\scriptstyle {i,j,\sigma}} (\tilde c_{i,\sigma}^{\dagger}
\tilde c_{j,\sigma}
+H.c.)+
J \sum_{\scriptstyle{i,j}}(S_i S_j - n_i n_j/4)+ \nonumber \\
\sum_{\scriptstyle{q}}((\omega(a_q^\dagger a_q +1/2)+  
g n_i(a_q^\dagger+a_{-q}))
\end{eqnarray}

At this stage the Hamiltonian operates over the Hilbert space of fermionic
and bosonic variables. Since the bosonic levels could be infinitely
populated, we must adopt some criteria to render finite the dimension of our
Hilbert space. In the calculation we include all the phonon modes (the zero
mode is excluded because it represents only a change in the total energy).
We populate each mode with the mimimum number of states for which we
converge the observable under consideration. We find that
15-phonon states for each mode is enough, at least for not very large
electron-phonon interaction.

For small frequencies and for $J<J_N$ the system has an abrupt transition
characterized by a rapid increases of the phonon occupation number $%
<n_{ph}> $ for each mode. Fig. 2a shows the occupacion number of the phonon $%
(\pi ,\pi )$ $n_{\pi ,\pi }$ as a function of $\lambda $ for different
values of $J$ and for $%
\omega =0.1$. When $J>J_N$ the transition is smoother and it takes place at
small $\lambda$ values. One important difference with the adiabatic case
is that here the transition depends on $\lambda $ even for $0<J<J_N$.
We have seen
that the static phonon calculation always gives rise to a localization
for $\lambda > \lambda_c$. The localization
is not obvious in the dynamical calculation, but the behaviour of the phonon
occupation as a function of $\lambda$ and the close agreement between
$\lambda_c$ obtained with both methods suggest that for $\omega =0.1$
the instablity is similar to that observed in the static case.
We  think that the adiabatic calculation is still valid
for small values of $\omega $.

For small $\lambda$ the phonon occupation is rather
small indicating that the system is formed by renormalized electron and
phonons.

When the frequency increases the system is more stable as it is shown in
Fig.2b for $\omega =1.0$. The transition towards a polaronic state is less
evident than that for smaller $\omega $. We  conclude that
the hole moves, as in a pure $t-J$ model, but with a bigger effective mass.

It is important to point out that the static calculation underestimates the
critical value of $\lambda $, which in our calculation seems to be large.
Thus the polaron occurs only for large $\lambda $ and small $\omega$
values .

We finish the section showing some results on the Hubbard-Holstein (HH)
model. It is well known that the Hubbard model could be mapped on the
$t-J$ model with $J=4t^2/U$ for $U/t\gg 1$. Moreover, for $U=0$ the HH
model has become the usual Holstein model. Therefore, for the HH model, it
is interesting to study the crossover from the small polaron of the
uncorrelated electron-phonon system to the one in the $t-J$ Holstein model.

The Hamiltonian of the HH model is:
\newpage
\begin{eqnarray}
H=-t\sum_{\scriptstyle {i,j,\sigma}} (c_{i,\sigma}^{\dagger}c_{j,\sigma}
+H.c.)+
U \sum_{\scriptstyle{i}}n_i n_j +\nonumber \\ 
\sum_{\scriptstyle{i}}(\frac{P_i^2}{2M}+K/2 u_i^2-\alpha u_i n_i)
\end{eqnarray}

Similarly to the $t-J$-H case we are going to study the HH model using
static and dynamic phonon approximations. The notation in the Hamiltonian
is standard. Note that we are studing the one hole doped system (3 electrons
in our cluster). This is different from the one particle problem previously
studied by different authors\cite{kabanov,alemanes}.

In Fig.3 we show the adiabatic calculation of a site deformation (the
biggest deformated one) as a function of $\lambda $ for different values of
$U$ (the analogue of Fig. 1 for the $t-J$-H model). In Fig. 4 we show the
occupation of the ($\pi,\pi$) mode (dynamic calculation).

For a very large $U$ ($U>U_c$ ,where $U_c$ is of order of $40$) the model
behaves like the $t-J$-H model.  That means: we obtain an abrupt transition
from a totally delocalized state to a full localized state. This transition
occurs
for $\lambda $ $\sim 2$ . This is exactly what happens for the $t-J$-H model
for $J<J_N$. When $U$ decreases $\lambda _c$
 also decreases and the transition is less abruted.
We found, like for the $t-J$ model case, that before going to the polaron
state the system has the
breathing-type state for $U<U_c$. Note the following different behaviour
between the $t-J$-H and HH model: in the former $\lambda _c$ always decreases
as $J$ increases showing the strong self-localization of the hole in the
antiferromagnetic backgroung. In the HH model case $\lambda _c$ goes to a
finite value when $U$ goes to zero, this is precisely the threshold for the
appearence of the polaronic band in the pure Holstein model.

For small $\omega $ and large U the average
of the phonon occupation presents an abrupt transition at a given value of $%
\lambda $ which is close to the $\lambda_c$ obtained using adiabatic 
approximation. Moreover       
, for small U and large $\omega$ the transition is smoother
and occurs at smaller $\lambda$. 
 Therefore,
stability of the system is reinforced when both $\omega $ and $U$ increase.

\section{Discussions and Conclusions}

In this paper we have studied the stability of correlated electronic systems
under the influence of a Holstein electron-phonon interaction. Both $t-J$
and Hubbard models have been analized using adiabiatic and dynamical phonon
approximations.

The important features found in the adiabatic calculation on the $t-J
$-H model are:

\begin{itemize}
\item  {\ Polaronic instability for the electron-phonon
couplings $\lambda > \lambda_c$.}

\item  {\ The critical value of the electron-phonon interaction decreases
as J increases. This is a consequence of the preexisting self-localization
for the pure $t-J$ model case.}

\item  {\ The critical electron-phonon coupling is
$J$- independent in the Nagaoka regime ($0.0<J<J_N$).}

\item  {In the Nagoka regime the transition towards a polaron is rather
abrupt while it is smoother when $J>J_N$.}
\end{itemize}

Treating the phonon dynamically we found that most of the results obtained
adiabatically are still valid for finite but small $\omega$ values.
A difference is that the value of $\lambda_c$ obtained in the dynamical
case is not independent of $J$ for the $0.0<J<J_N$ regime.

Increasing $\omega$, the transition towards the polaron state is less evident
and the system behaves like a liquid formed by renormalized electrons and
phonons.

It is important to point out that the adiabatic calculation understimates the
stability of the system. The system is more stable when $J$ decreases and $%
\omega$ increases.

We have also performed a calculation for the Hubbard model. In this model
most of conclusions made for the $t-J$ model were recovered for
large U. For
small U the stability of the system tends to that of the pure Holstein model
and the critical $\lambda$ goes to a finite value when U goes to zero.

When this paper has been finished we received a preprint of Wellein, Roeder
and
Fehske \cite{Fehske} where the Holstein Hubbard and $t-J$ Holstein models
were studied on finite cluster up to ten sites by an exact diagonalization
technique. The main results of this paper are consistent with the ones
obtained here. Moreover,
for one hole doped $t-J$-H model 
 they found evidence for the appeareance of a polaronic state
from a critical value of electron-phonon coupling. As this value is similar
to the one obtained in our paper, we believe that finite size effect are not
very important. We thank Dr. Fehske for sending us a copy of his paper prior
its publication.

We would like to thank J. Riera, M. Stachiotti and V. Kabanov 
for valuable discussions, and Fundaci\'on Antorchas
for financial support.

\newpage
\begin{center}
{\Huge {\bf Figure Captions:}}
\end{center}

\vskip 2.0cm 
{\large {\bf Figure 1:}} Displacement of one of the ions (the
one who's shift from the equilibrium position is maximum) obtained with the
iterative method descripted in the text for the $t-J$ Holstein model. 

\vskip 1.0cm 

{\large {\bf Figure 2:}} Occupation of the phonon of $k=(\pi,\pi)$ as
a function of $\lambda$, in the full quantum calculation including 15 state
for each phonon mode.a)$\omega = 0.1$; b) $\omega=1.$ 

\vskip 1.0cm

 {\large {\bf Figure 3:}} The same as fig. 1 but for the Hubbard-Holstein model 

\vskip 1.0cm 
{\large {\bf Figure 4:}} The same as fig. 2 but for the
Hubbard-Holstein model

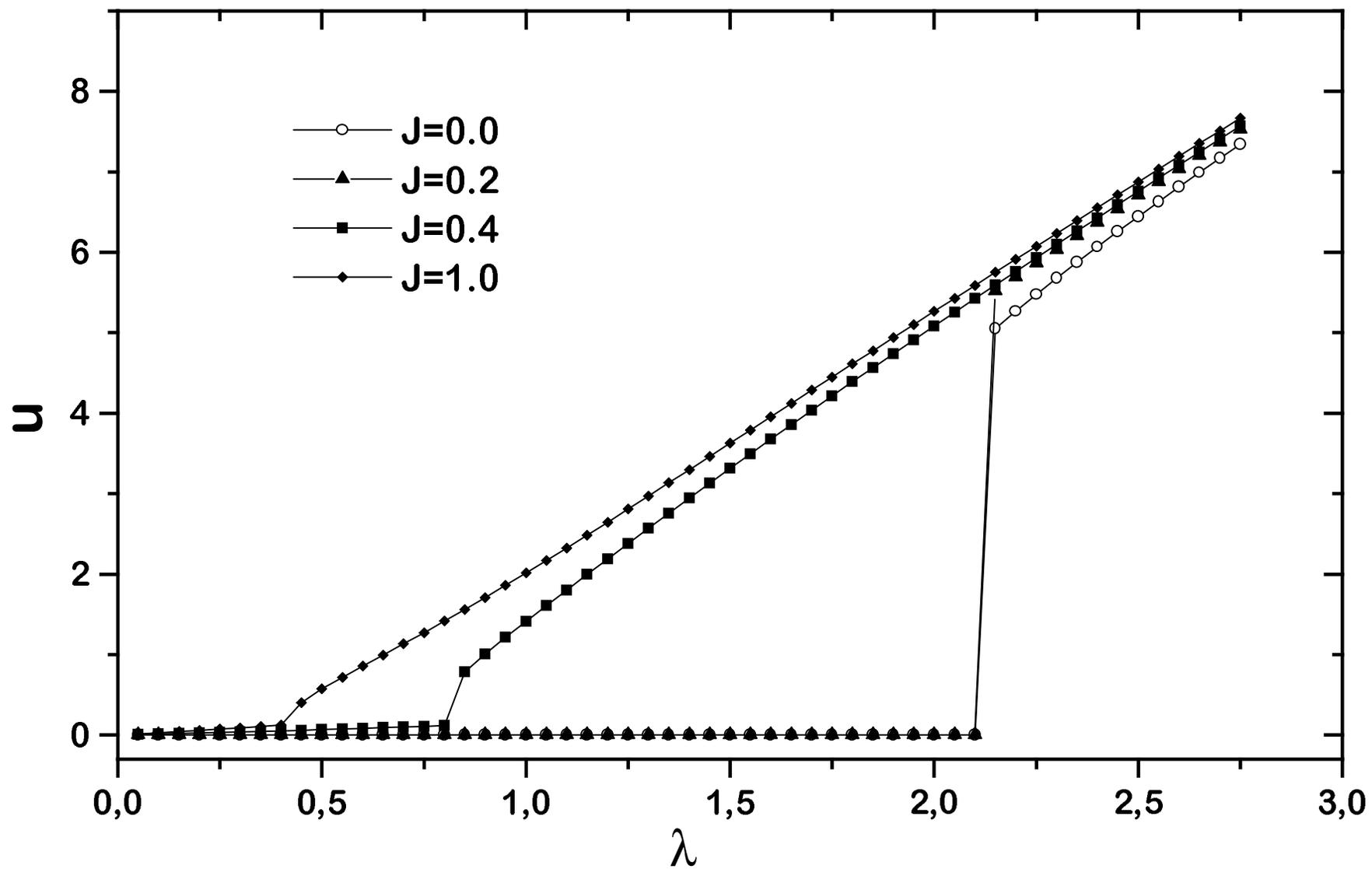
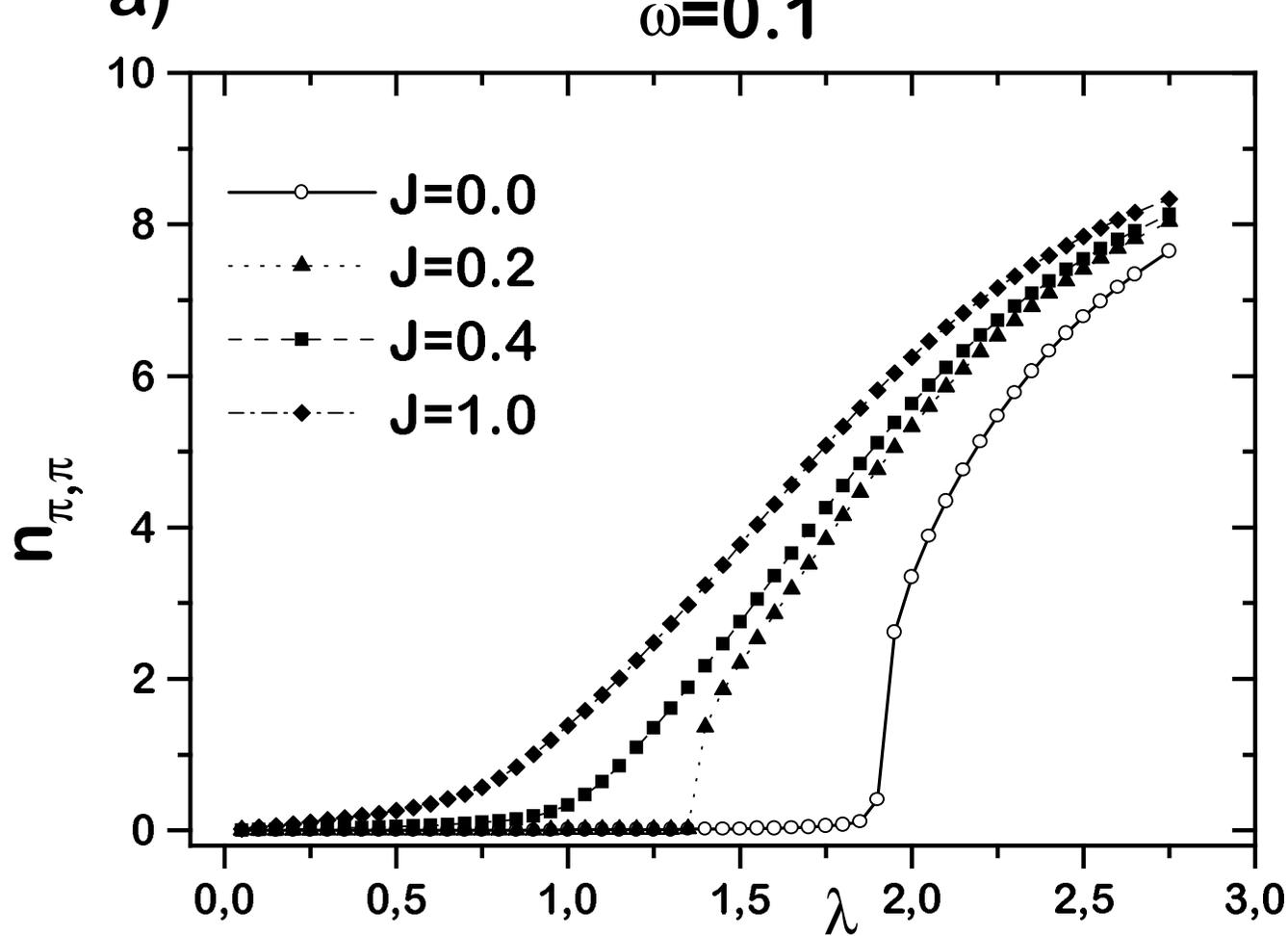
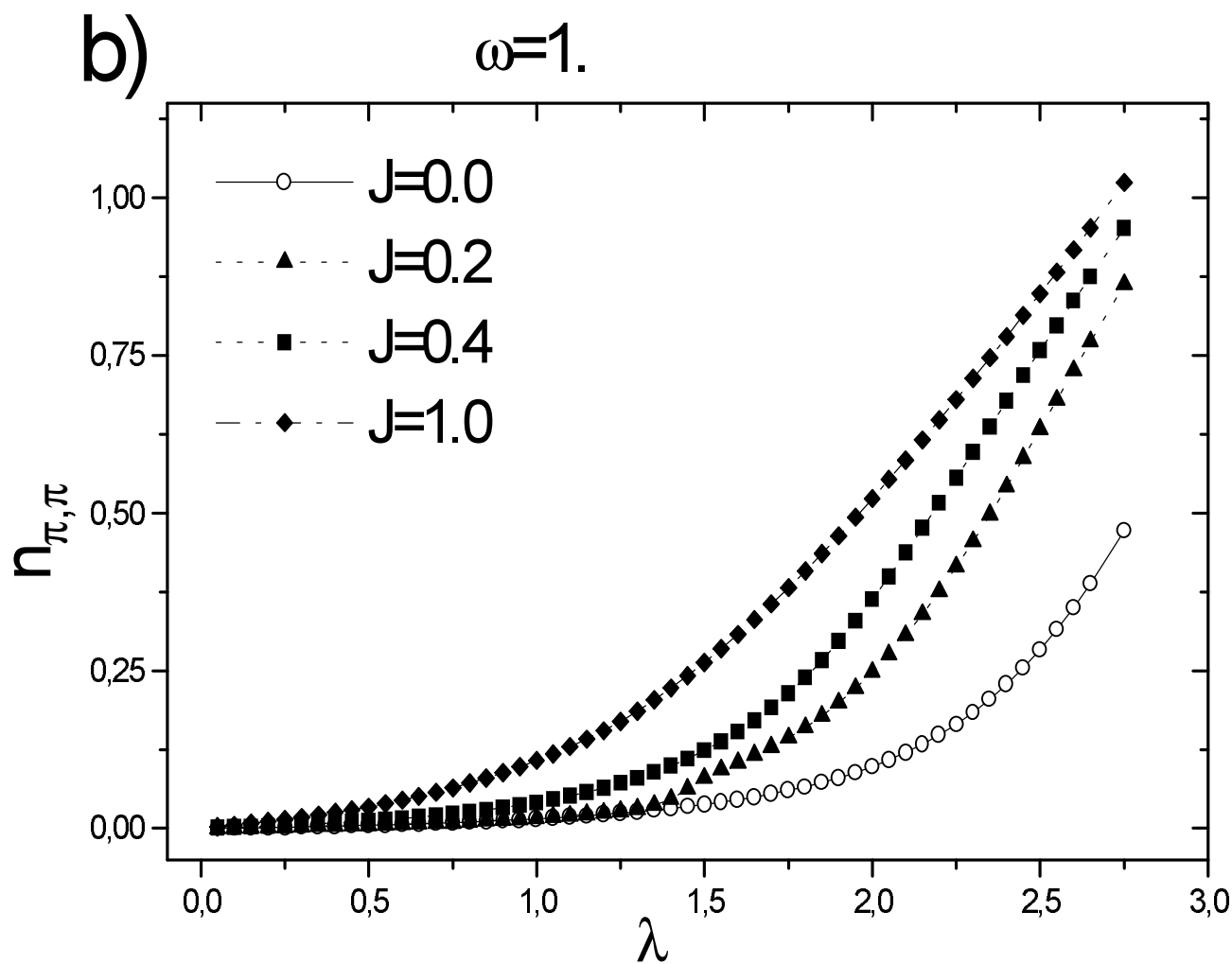
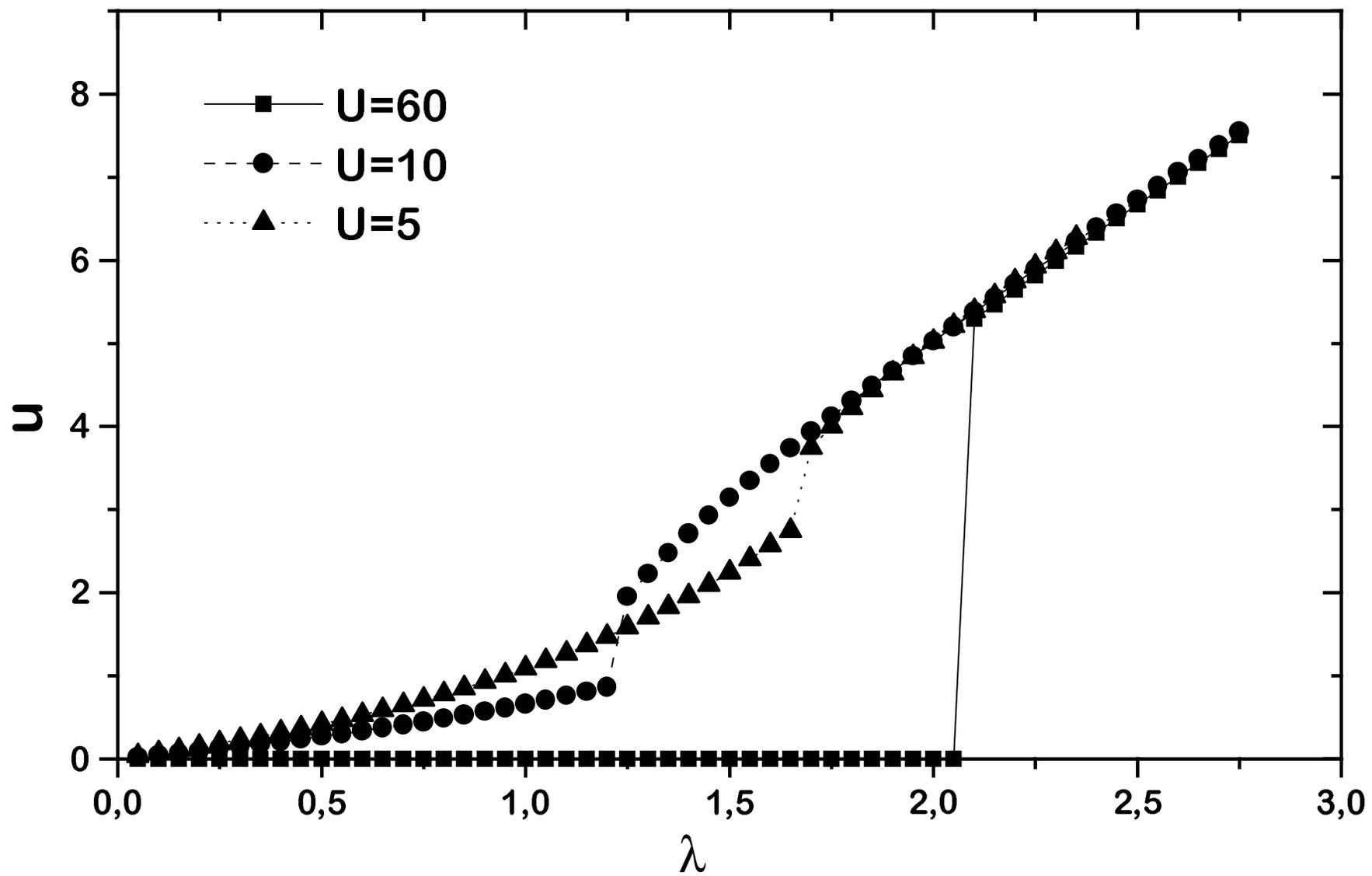
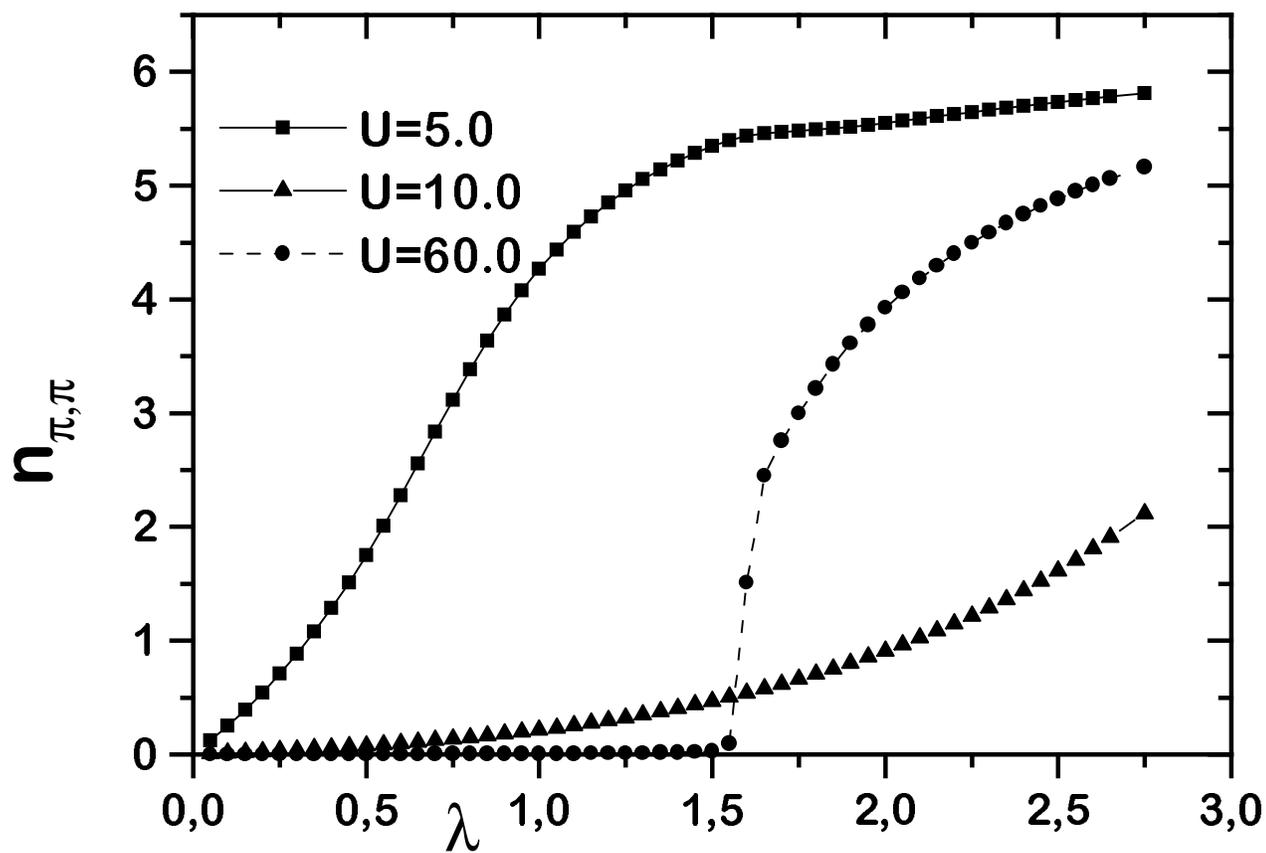
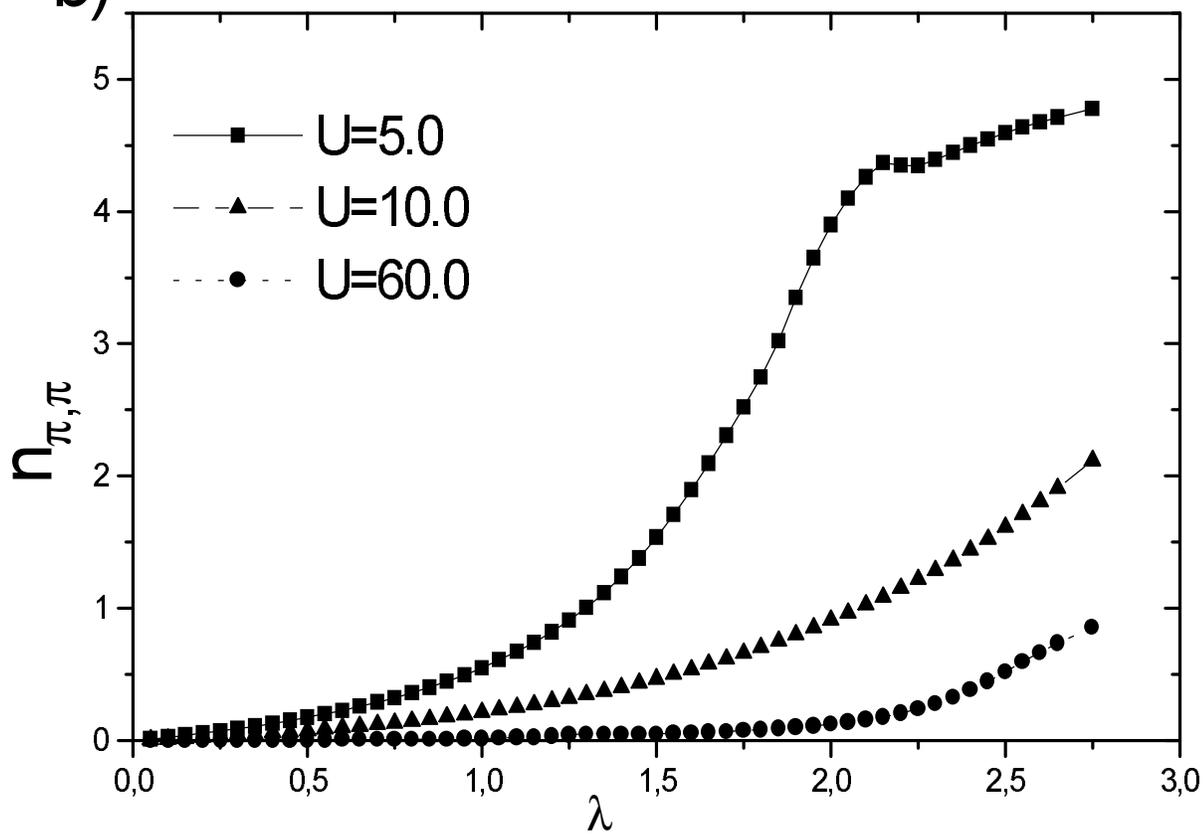
\end{document}